\documentclass[manuscript,nonacm]{acmart} 
\AtBeginDocument{%
  }

\setcopyright{acmlicensed}
\copyrightyear{2018}
\acmYear{2018}
\acmDOI{XXXXXXX.XXXXXXX}
\acmConference[Conference acronym 'XX]{Make sure to enter the correct
  conference title from your rights confirmation email}{June 03--05,
  2018}{Woodstock, NY}
\acmISBN{978-1-4503-XXXX-X/2018/06}
\usepackage{caption}
\usepackage{booktabs}   
\usepackage{multirow}   
\usepackage{subcaption}
\usepackage{graphicx}   
\usepackage{amsmath} 

\usepackage{array}       

\usepackage{caption}
\usepackage{algorithm}      
\usepackage{algpseudocode}  
\usepackage[justification=raggedright,singlelinecheck=false]{caption}
\usepackage{array} 
\begin{document}

\title{DiSCo: Making Absence Visible in Intelligent Summarization Interfaces}
 \author{Eran Fainman}
\orcid{0009-0004-3898-2086}
\affiliation{%
  \institution{University of Haifa}
  \city{Haifa}
  \country{Israel}}
\email{eranfineman@gmail.com}

\author{Hagit Ben Shoshan}
\email{hagitphd@gmail.com}
\orcid{0009-0007-5945-695X} 
\affiliation{%
  \institution{University of Haifa}
  \city{Haifa}
  \country{Israel}}

\author{Adir Solomon}
\orcid{}
\affiliation{%
  \institution{University of Haifa}
  \city{Haifa}
  \country{Israel}}
\email{asolomon@is.haifa.ac.il}

\author{Osnat Mokryn}
\orcid{0000-0002-1241-9015}
\affiliation{%
  \institution{University of Haifa}
  \city{Haifa}
  \country{Israel}}
\email{omokryn@is.haifa.ac.il}

\renewcommand{\shortauthors}{Fainman et al.}

\begin{abstract}
  Intelligent interfaces increasingly use large language models to summarize user-generated content, yet these summaries emphasize what is mentioned while overlooking what is missing. This presence bias can mislead users who rely on summaries to make decisions. We present Domain Informed Summarization through Contrast (DiSCo), an expectation-based computational approach that makes absences visible by comparing each entity’s content with domain topical expectations captured in reference distributions of aspects typically discussed in comparable accommodations. This comparison identifies aspects that are either unusually emphasized or missing relative to domain norms and integrates them into the generated text. In a user study across three accommodation domains, namely ski, beach, and city center, DiSCo summaries were rated as more detailed and useful for decision making than baseline large language model summaries, although slightly harder to read. The findings show that modeling expectations reduces presence bias and improves both transparency and decision support in intelligent summarization interfaces.
\end{abstract}

\ccsdesc[500]{Human-centered computing~User studies}
\ccsdesc[500]{Information systems~Expert systems}
\ccsdesc[500]{Computing methodologies~Artificial intelligence}
\keywords{Review Summarization, Absence, Expectations, Learning via surprisability, Missing commonalities}


\received{20 February 2007}
\received[revised]{12 March 2009}
\received[accepted]{5 June 2009}

\maketitle

\section{Introduction}

Consider a traveler evaluating beach resorts in Greece who encounters this summary:
\textit{``Guests consistently praised the hotel's warm, helpful staff and convivial atmosphere, noting a welcoming, cozy feel and comfortable beds. The location is also seen as pleasantly convenient. The restaurant and especially the breakfast drew strong praise for tasty, well-prepared meals.''}
The hotel has hundreds of reviews. The interface presents this summary as a decision aid, condensing hours of reading into a few sentences. Yet what is missing? Despite being classified as a \textit{beach resort}, there is not a single mention of beach quality, proximity, or views, which are all topics discussed in most comparable properties. For a user choosing among coastal accommodations, this absence is telling. It suggests either poor beach access or a distance from the waterfront that contradicts the label. In this case, the interface failed the user not by presenting false information but by \textit{presenting incomplete expectations}. 

Traditional review interfaces, even those powered by large language models (LLMs), surface only what reviewers explicitly mention, leaving critical gaps invisible to users making important decisions.
When users rely on AI-generated summaries, they implicitly assume the system presents a complete picture. Yet they cannot distinguish between ``not mentioned because unremarkable'' and ``not mentioned because problematic''. Absence is meaningful only against a reference frame of expectation~\cite{kahneman2011thinking, mumford2021absence}: we notice a missing email because we expected one, or a review lacking location remarks because comparable accommodations routinely discuss it. Existing summarization pipelines, however, are \textit{presence-driven}~\cite{nenkova2011automatic}: they identify what is most frequently mentioned and surface it to users, whether through extractive selection~\cite{liu2019text}, neural generation~\cite{chu2019meansum}, or LLM-based synthesis~\cite{zhang2024comprehensive}. Even recent opinion-summarization benchmarks~\cite{siledar2024one} evaluate only what is explicitly present, reinforcing intelligent systems’ bias toward visible information.

This limitation reflects a deeper cognitive asymmetry in human–AI interaction. The \textit{feature-positive effect}~\cite{newman1980feature} shows that people detect present cues more readily than absent ones and often overlook missing information unless explicitly guided~\cite{hearst1989backward}. Judgments of absence are slower, less confident, and strongly dependent on contextual support~\cite{coldren2000asymmetries, tversky1974judgment}. 
In cognitive terms, absence is perceived when an internal prediction is violated. That is, when what we observe \textit{diverges} from what we \textit{expect}~\cite{margoni2024violation}. Intelligent interfaces can leverage this principle by \textit{externalizing expectations}: if a system exposes what \textit{should} appear based on comparable entities, both absences and other deviations from expectation become perceptible~\cite{attfield2010sense}.

Detecting absence computationally requires constructing domain-level expectations against which each instance can be compared. Recent work in expectation-based text analysis formalizes this through population-level reference models. \textit{Learning via Surprisability} (LvS)~\cite{mokryn2025interpretable} builds such models by creating full-support reference distributions and quantifying how individual instances diverge from them. Building on this perspective, we develop \textbf{DiSCo}, \textit{Domain-informed Summarization through Contrast}, an expectation-based framework that quantifies how each accommodation’s topic distribution deviates from domain-level topical norms. Specifically, we identify aspects that are unusually frequent and domain-prevalent yet absent in a given accommodation, using these deviations to enrich AI-generated summaries with diagnostic signals.

Recently, the HIRO framework~\cite{hosking2024hierarchical} constructs a latent hierarchy over all reviews in the dataset, identifying the most informative nodes to guide review selection for each accommodation. In doing so, it generates summaries grounded in the most meaningful and representative opinions, while overlooking missing but commonly discussed topics. Similar to our approach, HIRO contextualizes opinion summaries through comparisons with other accommodations in the dataset. However, our method extends this comparative perspective by quantifying topic-level deviations from domain expectations, explicitly distinguishing between unexpected presences and missing domain-prevalent aspects. Moreover, rather than considering all reviews in the dataset, we focus on accommodation domains to achieve stronger alignment with user expectations. Using the \textit{Domain Topical Expectations} as a reference, DiSCo measures how each accommodation’s topic distribution diverges from expectations derived from comparable entities (e.g., beach resorts in Greece and Spain). This decomposition isolates aspects that are unusually frequent, and domain-prevalent aspects that are absent. We then integrate these deviation profiles into LLM-based summary generation, allowing the interface to highlight both expected and missing information. The resulting \textit{DiSCo summaries} embed domain expectations directly into the generated text, revealing existing gaps in reviews  and improving alignment with user expectations.

To evaluate how this approach affects users’ perception of AI-generated summaries, we conducted a user study across three accommodation domains: Ski, Beach, and City-center accommodations. For each accommodation, participants viewed two versions of an AI-generated summary: a baseline LLM summary reflecting only topics explicitly mentioned in the reviews of the accommodation, and a summary generated using DiSCo that highlighted topics unusually present or missing relative to the \textit{Domain Topical Expectations}. Participants rated each summary along five evaluation dimensions: relevance, detail and specificity, helpfulness, decision support, and ease of understanding. Across all domains, DiSCo summaries were perceived as more detailed, helpful, and decision-oriented, with a modest reduction in ease of understanding. These findings indicate that explicitly representing deviations from expectation, and particularly absences, enhances the informativeness and practical value of intelligent review interfaces.

\noindent \textbf{Our contributions are as follows:}
\begin{itemize}
    \item We identify a fundamental limitation in intelligent review interfaces: presence-driven summaries obscure diagnostic gaps that arise from missing information.
    \item We introduce \textbf{DiSCo}, \textit{Domain-informed Summarization through Contrast}, a framework that constructs domain-level topical references and detects which aspects are uniquely emphasized or absent in each accommodation’s reviews relative to similar establishments.
    \item We integrate expectation-based analysis with LLM-based summarization to produce absence-aware interfaces that externalize what \textit{should be present} as well as what is.
    \item Through a multi-domain user study, we demonstrate that making absences explicit improves perceived informativeness and decision support compared with strong LLM baselines.
\end{itemize}

\section{Related Work}

In this section, we first describe the recent advances in review summarization, followed by a discussion of \textit{Absence}—a critical yet often overlooked signal in reviews.

\subsection{Review Summarization}

Early studies on review summarization relied on basic aspects and sentiment classification~\cite{hu2004mining,popescu2005extracting}. These works extracted key insights using rule-based aspect-extraction methods, often identifying frequent noun phrases as indicators of product features. However, because aggregating aspects and sentiments captures only fragments of the overall user experience,~\citet{carenini2006multi} shifted toward generating free-text summaries that combine multiple reviewer perspectives. ~\citet{lerman2009sentiment} aimed to improve coverage and coherence by selecting representative sentences from large review corpora and identifying the most salient viewpoints across diverse collections.

Subsequent approaches explored unsupervised techniques for discovering aspects and their associated sentiments, including clustering, topic modeling, and keyword-based methods~\cite{levi2012finding,chen2016clustering,nazir2020issues,mokryn2024consumer}. Summaries were then produced for each aspect by selecting representative comments using feature scoring or, later, deep learning approaches. The SemEval shared tasks on aspect-based sentiment analysis~\cite{pontiki2014semeval,pontiki2016semeval} established standardized benchmarks that significantly advanced both aspect extraction and sentiment classification. However, these benchmarks evaluate only explicitly mentioned aspects, thereby reinforcing the field’s focus on surface-level expressions of opinion. Subsequent neural models, such as MeanSum~\cite{chu2019meansum}, the extractive transformer approach ~\cite{angelidis2021extractive}, and controllable summarization frameworks~\cite{isonuma2019controllable}, introduced greater flexibility but continued to depend primarily on explicitly stated information.

The emergence of LLMs has transformed review summarization by enabling more abstract reasoning and flexible semantic understanding. Rather than operating through predefined features or supervised training data, LLMs can generate coherent summaries in zero-shot or few-shot settings by interpreting reviews in context. ~\citet{zhang2023benchmarking} demonstrated that instruction tuning—how the task is framed in the prompt—plays a greater role than model size in determining summarization quality, while~\citet{liu2024benchmarking} showed that even state-of-the-art models still struggle with maintaining factual consistency and controllable coverage across multiple evaluation dimensions. 

However, while LLMs now enable flexible, zero-shot summarization across diverse domains, many remaining challenges remain, such as factual consistency, domain alignment, and alignment with human evaluations~\cite{zhang2025systematic}. Recent hybrid frameworks combine human feedback and structured intermediate representations (e.g., topic models, keyphrase scaffolds) with generative modeling~\cite{liu2023pre}. For example, the LimTopic framework~\cite{limtopic2024} combines BERTopic with GPT-based summarization to generate coherent topic titles and summaries that better capture latent semantic structure.

Despite these advances, current approaches focus solely on information explicitly present in the data. They do not account for absent but expected information—topics or aspects that typically appear in comparable contexts but are systematically missing in a given dataset or review set. Such absences, when interpreted against population-level expectations, can carry diagnostic meaning, revealing unmet user expectations, overlooked experiences, or service gaps that presence-based methods fail to detect.

\subsection{Human-centered summaries' evaluations}
LLM-based summarizations can be inconsistent (or even hallucinate), and lack specific domain alignments~\cite{zhang2025systematic}. To make opinion summarization more transparent and human expectation-aware, recent hybrid frameworks combine human feedback and structured intermediate representations (e.g., topic models, keyphrase scaffolds) with generative modeling~\cite{liu2023pre}. Such approaches aim to produce summaries that are not only fluent but also \textit{faithful, interpretable, and contextually grounded}.

Early evaluation methods compare machine-generated summaries to human references by counting the number of words or short phrases that overlap. Metrics such as ROUGE and BLEU remain widely used for this purpose, but they measure only surface similarity and do not capture whether the information is factually correct or complete~\cite{zhang2025systematic}. More recent embedding-based measures, such as BERTScore and BLEURT, use neural models to assess semantic similarity rather than exact word-for-word matches. While they better reflect semantic quality, their scores are harder to interpret and can still reward fluent yet inaccurate summaries.

Recent LLM-based evaluators address these limitations by acting as judges.  
Frameworks such as GPTScore and G-Eval employ structured prompts and chain-of-thought reasoning to assess multi-dimensional quality criteria, showing strong alignment with human ratings~\cite{zhang2025systematic}.  
The SUMMEVAL-OP benchmark is an LLM-as-judge for opinion summarization~\cite{siledar2024one} that uses LLM evaluators to assess the summaries over seven dimensions: \textit{fluency, coherence, relevance, faithfulness, aspect coverage, sentiment consistency, and specificity}.  The goal is to capture a different facet of summary quality, from grammatical smoothness and logical flow to factual accuracy, breadth of topics, correctness of sentiment tone, and the use of concrete details rather than vague statements. Human annotators rated summaries on a five-point scale for each dimension, and LLMs were prompted with the same criteria to assign their own scores, achieving up to 0.70 Spearman correlation with human judgments.  

An interesting recent critique of human evaluations as gold standards analyzes how human preference scores distort evaluation~\cite{hosking2023human}.  
The authors show that single ``quality'' ratings under-represent factuality and consistency, are biased by surface fluency, and are strongly confounded by \textit{assertiveness}. That is, \textit{annotators systematically rate confident but incorrect outputs as more factual and higher quality}.  
They conclude that reward models trained on such feedback risk amplifying assertiveness and style biases rather than genuine quality improvements.

The HIRO framework~\cite{hosking2024hierarchical} introduced a comprehensive evaluation methodology for opinion summarization that combines human preference judgments with automatic entailment-based metrics. Human annotators compared HIRO-generated summaries with baseline outputs using pairwise preference judgments. Each pair was evaluated by three annotators along four qualitative dimensions: \textit{accuracy} (faithfulness to the opinions and facts expressed in the reviews), \textit{detail} (amount of concrete, informative content), \textit{coherence} (logical flow and readability), and \textit{overall quality} (a holistic preference combining faithfulness, informativeness, and fluency). Annotators viewed the original reviews and selected which summary better captured their content and tone; final decisions were aggregated by majority vote. These judgments emphasized that perceived quality depends more on factual coverage and specificity than on stylistic fluency.

Beyond human evaluation, HIRO also proposed a reference-free, entailment-based scheme to assess how well summaries reflect underlying reviews through three complementary metrics: \textit{Prevalence}, \textit{Genericness}, and \textit{Specificity-Adjusted Prevalence (SAP)}. \textit{Prevalence} measures how representative each statement is of the input reviews, \textit{Genericness} captures how often the same statement appears across unrelated entities, and \textit{SAP} combines the two ($\text{SAP} = \text{Prevalence} - 0.5 \times \text{Genericness}$), rewarding statements that are both well supported and specific to the summarized entity. Together, these metrics encourage summaries that are distinctive yet grounded in the source material.

Similar to HIRO, our approach contextualizes opinion summaries relative to comparable entities, such as generating an accommodation summary while considering reviews of similar accommodations. However, we extend this comparative perspective by quantifying topic-level deviations from domain expectations using LvS. This method not only captures what is unusually emphasized but also reveals notable absences—an aspect of distinctiveness that purely presence-based or entailment-based models overlook.

\subsection{Presence Bias and the Importance of Absence}

Presence bias is the systematic tendency to detect, learn, and reason from things that are present more readily than from things that are absent. 
In learning and judgment, this asymmetry is known as the \textit{feature-positive effect (FPE)}: people process and associate present cues more fluently than missing ones, easily learning ``A and B'' but struggling with ``A and not B''~\cite{newman1980feature}. 
Evidence for this bias appears early in development, as infants pay more attention to added than to removed features, revealing a deep cognitive asymmetry~\cite{coldren2000asymmetries}. 
Kahneman similarly observes that the human mind constructs coherent narratives from available cues while often overlooking what is missing unless it is made explicitly salient~\cite{kahneman2011thinking}. 
These cognitive tendencies help explain why current computational and generative approaches to summarization focus almost exclusively on information that is explicitly \textit{present} in text. 
Models are optimized to detect and reproduce salient linguistic signals rather than to infer what is \textit{missing but expected}. 
For instance, a model trained on review data may easily learn to highlight repeated mentions of ``great location'' or ``friendly staff'' but it will not note that ``breakfast'' or ``air conditioning'' are never discussed, despite such omissions being meaningful in context. 

In philosophy, psychology, and cognition, absence refers to missing but expected elements in a dataset or discourse, features whose absence carries informational value because it violates shared expectations~\cite{tversky1974judgment,newman1980feature,mumford2021absence,margoni2024violation}. 
We notice absence only against a reference frame of what should have been present: an email that never arrives, a section missing from a scientific paper, or a review that omits any mention of the location. 
Philosophical and cognitive theories describe this as a hybrid phenomenon: our perception registers what exists, while cognition compares it with internal expectations, and the discrepancy gives rise to the experience of absence~\cite{mumford2021absence,farennikova2013seeing}. 

In visual reasoning, ~\citet{attfield2010sense} extend this principle to analytic contexts, describing sense-making as an iterative loop between data foraging and updating explanatory hypotheses.  Effective visual analytics systems, they argue, are those that make expectations and discrepancies explicit, allowing users to recognize both anomalies and missing information.

\subsection{Identifying Absences in Texts}
Modern machine learning achieves remarkable success on high-dimensional data, but often operates as a black box, providing little insight into which features drive its predictions~\cite{rudin2019stop}.
In critical applications such as medical diagnosis, scientific discovery, and security, understanding what makes data distinctive is as important as achieving high predictive accuracy.
However, quantitative data analysis has traditionally assumed that meaningful patterns arise solely from correlations between \textit{present} features~\cite{vapnik2013nature}.
Traditional text summarization and sentiment analysis pipelines seldom account for missing or underrepresented elements, implicitly assuming that the most informative cues are those explicitly mentioned~\cite{nenkova2011automatic}.
Extractive and abstractive summarization methods, whether based on frequency statistics, neural attention mechanisms, or transformer architectures, emphasize \textit{presence salience}, identifying sentences or tokens that best represent the dominant content of the corpus~\cite{rush2015neural,liu2019text}.
Even with the advent of LLMs, which can flexibly integrate contextual and semantic information through in-context learning, model attention remains guided by visible tokens and surface-level co-occurrences rather than by what is conspicuously absent~\cite{zhang2024comprehensive}.

Recent approaches to text analysis and summarization increasingly recognize that meaningful interpretation depends not only on what is present in the data but also on what is missing.
For example, Latent Personal Analysis (LPA)~\cite{mokryn2021domain} and LvS~\cite{mokryn2025interpretable} both construct \textit{expectation references}—population-level models that represent collective regularities against which individual instances are compared.
The two frameworks differ in how these references are constructed, yet they share the goal of identifying deviation from the expected norm as a meaningful analytic signal. 

LvS represents each textual instance as a discrete probability distribution over a fixed feature space, such as normalized word frequencies, topic proportions, or sentiment-related lexical categories. A population reference distribution is constructed by aggregating these distributions across the corpus, forming an explicit statistical model of collective expectations. For each instance, LvS computes the divergence between its feature distribution and the population reference using Jensen–Shannon Divergence, which remains well defined for sparse and non-overlapping distributions. The divergence is decomposed at the feature level, allowing attribution of deviation to specific features. Features whose observed frequencies exceed the reference correspond to unexpected presences, while features that are prevalent in the reference but underrepresented or absent in the instance are identified as missing commonalities. The resulting feature-level divergence profile constitutes a surprisal signature that characterizes how the text deviates from population-level expectations.

The LPA/LVS frameworks are conceptually aligned with Bayesian and surprisal-based models of cognition, which interpret learning as the continual adjustment of expectations in light of prediction errors.
Unlike probabilistic text models that rely primarily on token likelihoods, LvS and LPA explicitly quantify how each instance diverges from the population reference, decomposing this divergence into interpretable components that expose both unexpected presences and missing commonalities, popular prevalent features (and hence globally expected) that are absent from an instance.
This decomposition reveals which expected topics, sentiments, or lexical fields fail to appear, making absence itself an informative and measurable property of textual meaning.

In the context of accommodation reviews, this means comparing the distribution of topics and their associated sentiments for one accommodation against the aggregated distribution across many comparable accommodations. 
A high divergence on a specific feature indicates an \textit{unusual emphasis}. For example, if guests at one accommodation mention ``noise'' or ``air conditioning'' far more often, or express unusually strong opinions about them. 
Conversely, a feature that is common elsewhere but rarely mentioned for a specific accommodation represents a \textit{notable absence}, such as when most accommodations receive many comments about ``breakfast\_positive, cleanliness\_positive'', but the target accommodation does not. By quantifying both unexpected presences and absences, the method reveals how each accommodation differs from collective guest expectations, capturing what stands out and what is notably absent in its feedback.

\section{Data Preparation}

\subsection{DiSCo Summary Generation Pipeline}

\begin{figure}[ht]
\centering
\includegraphics[width=\linewidth]{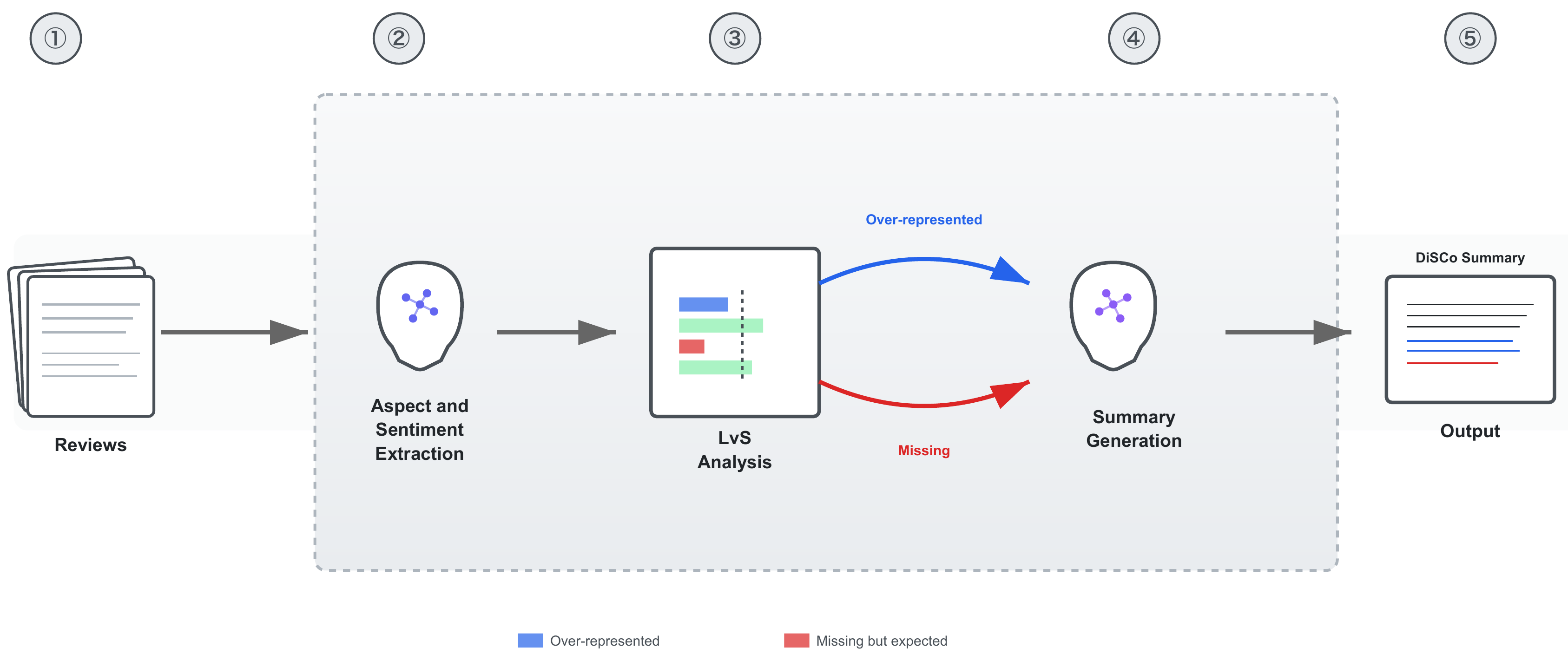}
\caption{%
\textbf{DiSCo pipeline.}
The process of Domain-informed Summarization through Contrast (DiSCo) integrates
expectation-based analysis with LLM-based summarization.
(1) User reviews are collected for all accommodations in a predefined area.
(2) Aspect and sentiment extraction identifies topics and their polarity.
(3) LvS computes divergences between each accommodation’s topic distribution and domain-level topical expectations,
revealing over-represented and missing (expected but unmentioned) aspects.
(4) These deviations are incorporated into a structured LLM prompt for
summary generation.
(5) The resulting DiSCo summaries explicitly surface both salient and absent information relative to domain norms.%
}
\label{fig:disco_pipeline}
\end{figure}

Figure~\ref{fig:disco_pipeline} depicts the \textbf{DiSCo} pipeline. 
\noindent \textbf{1. Data collection} We built our study on the Booking.com reviews dataset, downloaded to a dedicated server from the official repository\footnote{\url{https://github.com/bookingcom/ml-dataset-reviews}}. From this source, we derived three distinct domains, each corresponding to a different type of tourism context: 
(1) \textit{Ski}, consisting of accommodations in France and Austria labeled as ski resorts; (2) \textit{Beach}, consisting of accommodations in Greece and Spain labeled as beach resorts; and (3) \textit{City center}, consisting of accommodations in Spain and Greece labeled as city-center accommodations. Our data, prompts, together with LLM generation configuration and code, are available in our code 
repository\footnote{\url{https://anonymous.4open.science/r/lvs-summarization-C357/README.md}}.

\noindent \textbf{2. LLM-based Aspect and Sentiment Extraction.} Reviews from each accommodation are processed to identify topical aspects and their associated sentiments using an LLM-based extractor. Following prior studies on accommodations review analysis~\cite{sann2020understanding,guo2017mining,gerdt2019relationship,hu2019hotel}, we compiled a list of 138 topics organized in a two-level hierarchy (see Appendix~\ref{appendix:topics}). To analyze review aspects and their associated sentiments, we adopted an LLM-based approach inspired by \citet{jeong2024aspect}. Specifically, we used the \textit{GPT-5-mini} model\footnote{\url{https://openai.com/index/gpt-5-system-card/}}
 with a carefully designed system and user prompts, as detailed in Appendix~\ref{appendix:absa_prompts}.

To assess whether LLM-based aspect-sentiment extraction provides sufficiently reliable signals for downstream comparative analysis, we conducted a lightweight validation. We randomly sampled 150 reviews across the three domains and compared the outputs of our LLM-based extractor using the GPT-5-mini model against annotations produced by two strong LLMs (Claude-3.7-Sonnet\footnote{https://www.anthropic.com/news/claude-3-7-sonnet} and Gemini-2.5-pro\footnote{https://docs.cloud.google.com/vertex-ai/generative-ai/docs/models/gemini/2-5-pro}), used as independent reference annotators. Aspect identification achieved a micro-F1 score of 0.80 with Cohen's $\kappa = 0.61$, and sentiment polarity accuracy was 0.99. These agreement levels are consistent with prior work showing that LLMs can serve as reliable annotators for topic and sentiment-related labeling tasks, particularly for population-level and comparative analyses rather than fine-grained gold annotation \citep{hellwig2025exploring,gilardi2023chatgpt,artstein2008intercoder}. 

We note that aspect-sentiment extraction is used here as an intermediate representation to construct domain-level expectations rather than as a standalone prediction task, and is therefore not treated as a primary contribution of this work.

\begin{figure}[ht]
    \centering
    \includegraphics[width=0.75\linewidth]{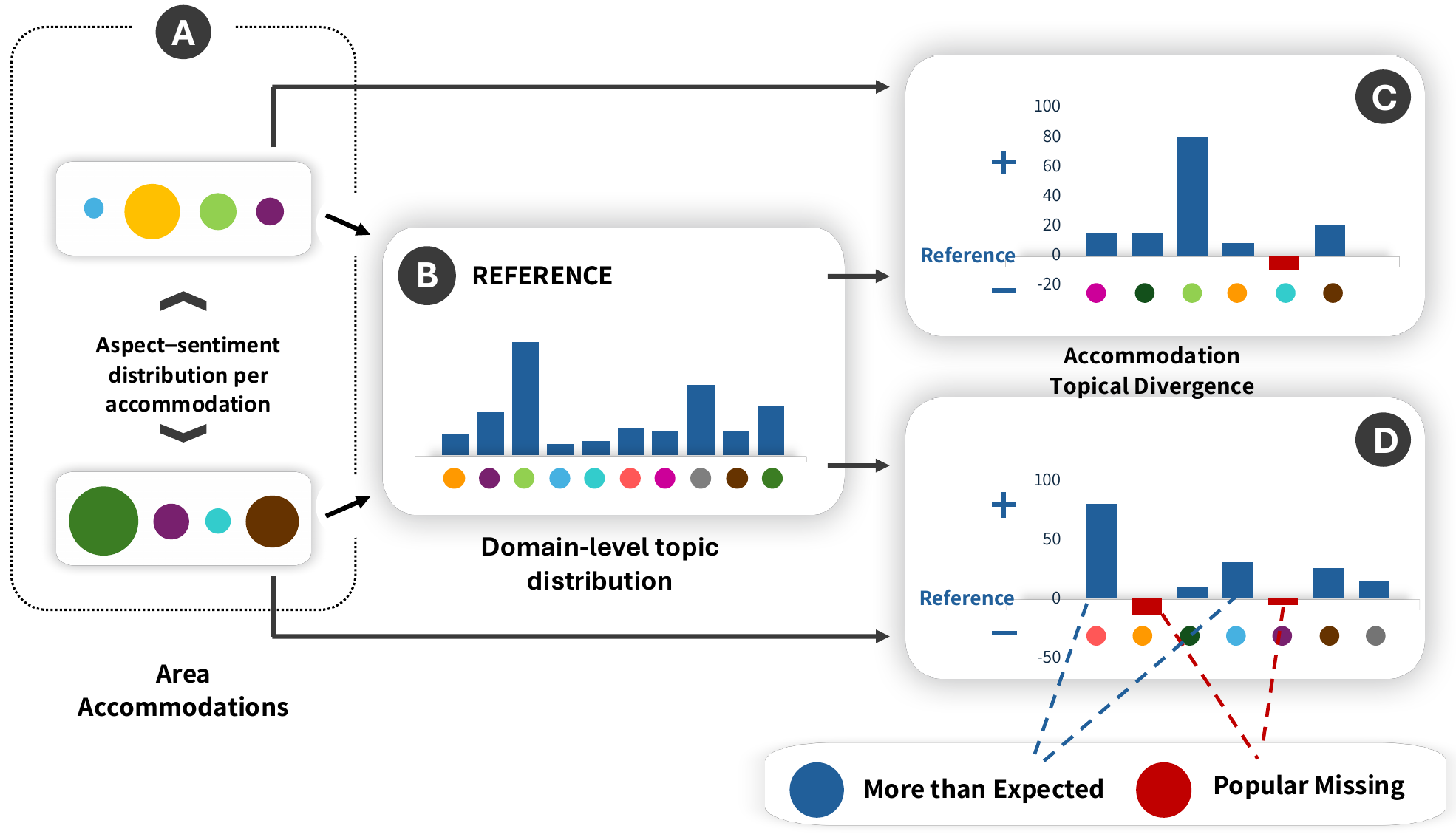}
    \caption{LvS-based analysis framework for identifying accommodation-specific topic deviations from reference distribution~\cite{mokryn2025interpretable}. (A) Input data: topic distributions within individual accommodations, where colored circles represent distinct topics and circle size indicates topic prevalence. These were identified in phase 2 of DiSCo. (B) Aggregate reference distribution computed across all accommodations in the corpus. (C--D) Accommodation-specific deviation analysis showing topics appearing in surplus (blue bars, above reference baseline) or absent (red bars, below reference baseline). Blue annotations indicate topics exceeding expected frequency, and red annotations indicate popular topics that are underrepresented or absent in the focal accommodation.}
    \label{fig:lvs}
\end{figure}

\begin{figure}[h]
\centering
\begin{minipage}{0.49\textwidth}
    \centering
\includegraphics[width=\linewidth]{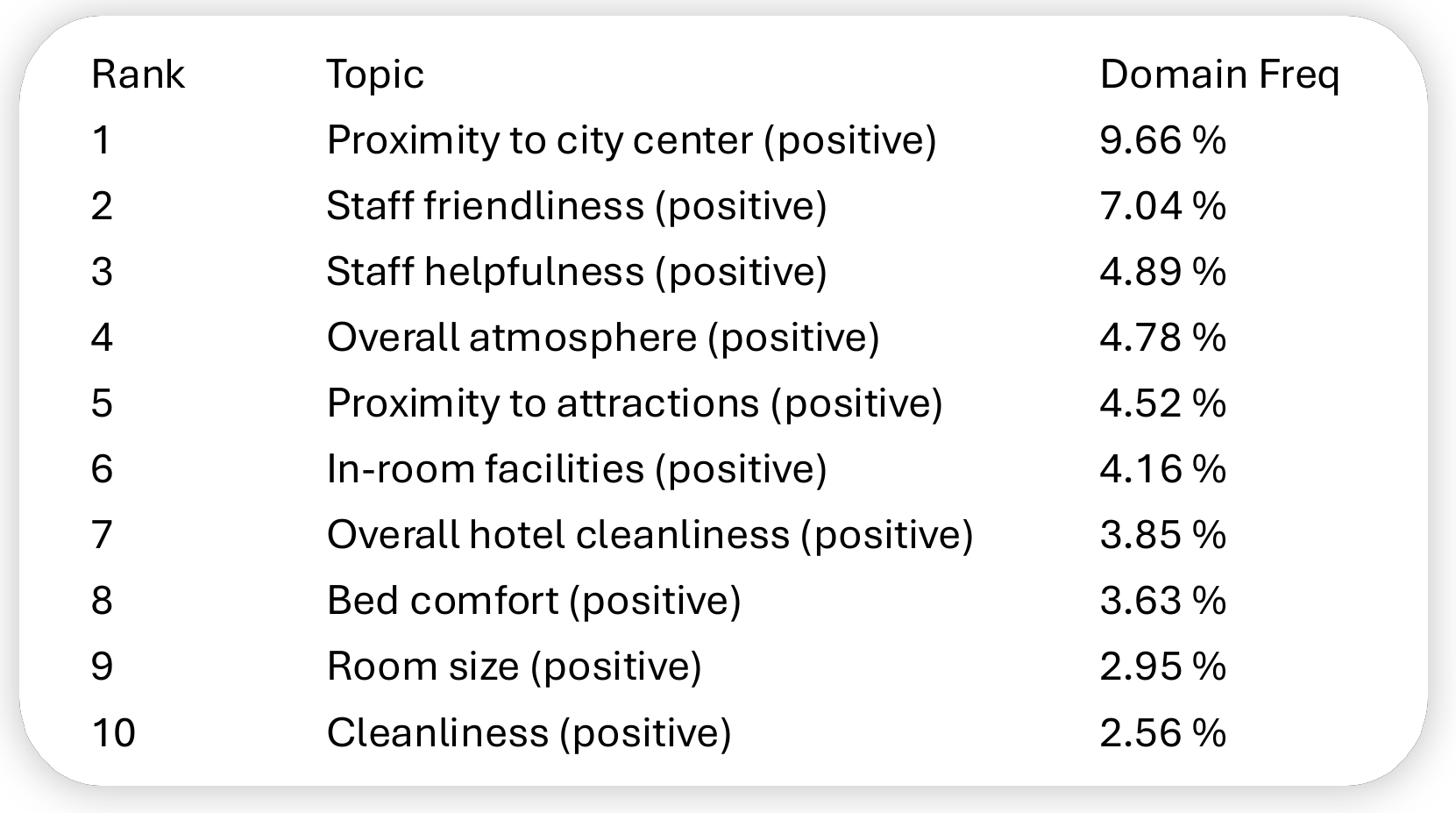}
    \caption*{\centering(a) City-center domain}
\end{minipage}
\hfill
\begin{minipage}{0.48\textwidth}
    \centering
    \includegraphics[width=.97\linewidth]{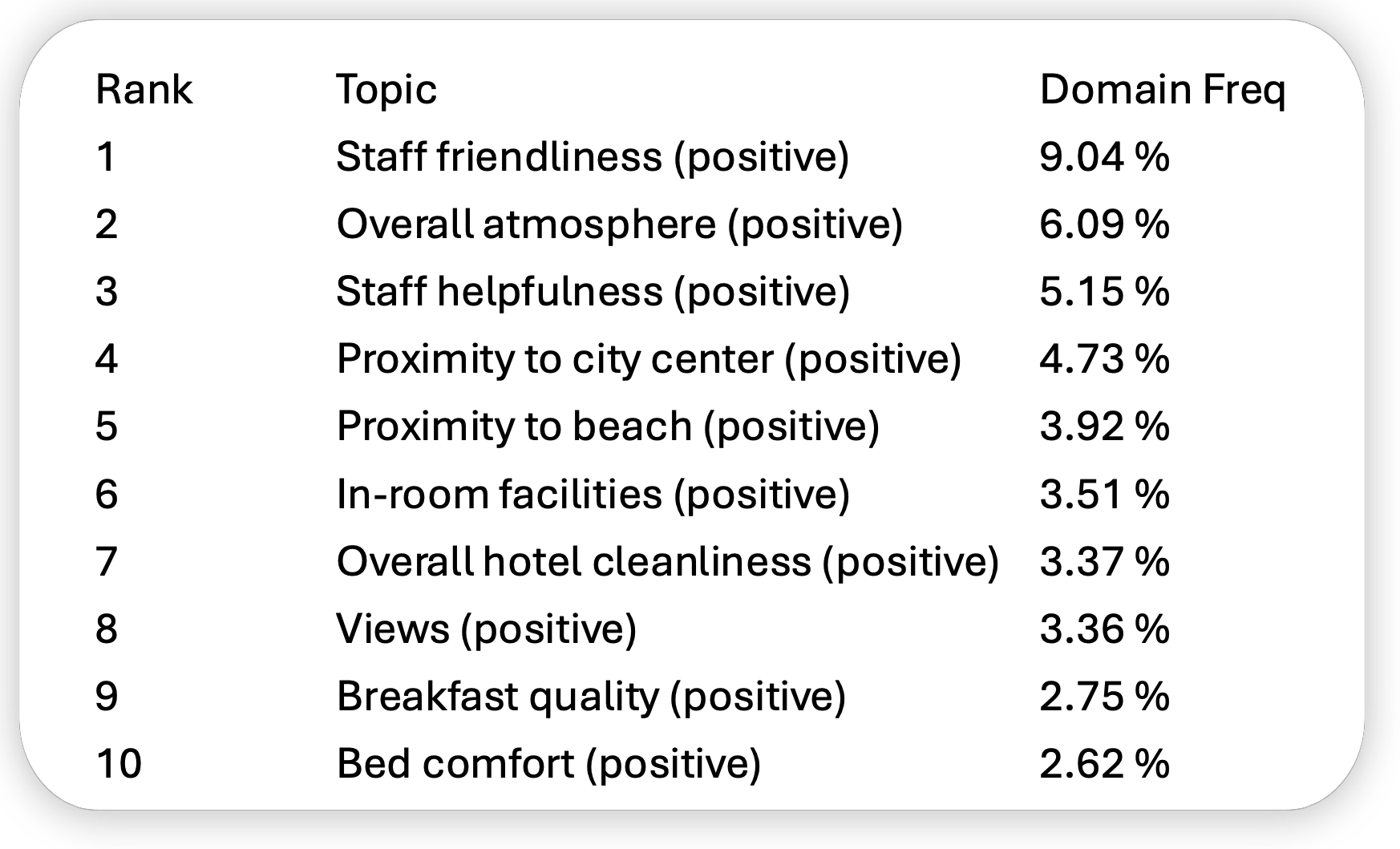}
    \caption*{\centering(b) Beach domain}
\end{minipage}
\caption{Top aspects in the City-center and Beach domains. Each bar represents the relative global weight of topics aggregated from all accommodations within the respective domain. The most prevalent aspects, such as \textit{staff friendliness}, \textit{location proximity}, and \textit{cleanliness}, define the domain’s typical focus and collectively form its domain topical expectations. These distributions represent what users commonly discuss and thus what others expect to appear in reviews for similar accommodations.}
\label{fig:domain_expectations}
\end{figure}

\begin{figure}[h]
\centering
\includegraphics[width=0.8\linewidth]{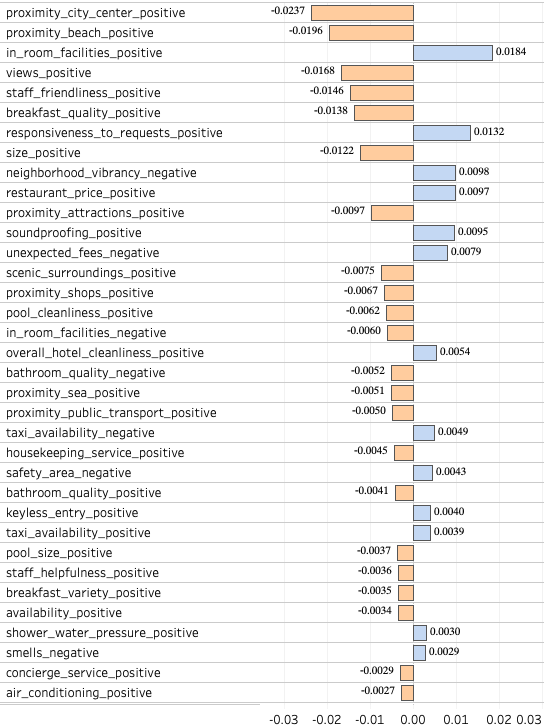}
\caption{\centering Example of presence–absence topic analysis for a single accommodation. 
Topics that appear more frequently than the domain average 
are shown on the right, while domain popular topics that are missing or hardly mentioned are shown on the left. Positive values indicate over-representation, and negative values indicate under-representation or absence.}
\label{fig:lvs_example}
\end{figure}

 \noindent \textbf{3. Identifying an accommodation's unique aspects profile}.  To create the accommodation's unique topics, either in surplus or domain-popular topics that are not mentioned for this accommodation, we employ a computational expectation-based method, LvS. It is operationalized to create \textit{Domain Topical Expectations} for each area, and then within each area an \textit{Accommodation Topical Divergence} for each accommodation, as follows.  
\begin{description}
\item[Domain Topical Expectations] We construct \textit{Domain Topical Expectations} that represent the aspects most commonly discussed within each accommodation domain (e.g., beach resorts, ski lodges, city-center hotels). We aggregate all reviews of all accommodations in that domain. Whenever a sentence discussed a topic with a specific sentiment from the list constructed in phase 2, the corresponding topic–sentiment tuple’s frequency is incremented. The resulting list of tuples, along with their normalized frequencies, represents the domain’s topical distribution, i.e., our \textit{domain expectations} as illustrated in Figure~\ref{fig:lvs}, Panel~A, and referred to as the \textit{reference}.
Figure~\ref{fig:domain_expectations} illustrates examples of \textit{Domain Topical Expectations} derived from the City-center and Beach domains. Each domain exhibits a distinct topical signature: city-center accommodations emphasize proximity, staff friendliness, and cleanliness, while beach resorts highlight beach access, views, and atmosphere. 
\item[Accommodation Topical Divergence] For each accommodation, we then measure how its topic distribution diverges from the corresponding Domain Topical Expectations. This step identifies aspects that are \textit{over-represented} (mentioned more than expected) and those that are \textit{absent but expected} (commonly discussed in the domain yet missing from the accommodation’s reviews). For each accommodation, we compute deviations from this reference by calculating the difference between observed and expected topic frequencies. Figure~\ref{fig:lvs} Panels C and D demonstrate this comparison for two representative accommodations, where positive deviations (blue bars) indicate topics appearing with greater-than-expected frequency, while negative deviations (red bars) identify topics that are underrepresented or absent despite being prevalent in the broader corpus. This approach enables the identification of distinctive topical emphases and notable content gaps for each property.
Figure~\ref{fig:lvs_example} shows an example of this analysis. Bars to the right indicate topics that appear more frequently for a given accommodation than in the domain as a whole, while bars to the left indicate topics that appear less frequently. For example, this accommodation has above-average mentions of \textit{`in room facilities positive'} and \textit{`responsiveness to requests positive'}, while mentions of \textit{`proximity city center positive'} and \textit{`views positive'}, which are popular in the domain, are below average or missing.
\end{description}

\noindent \textbf{4. DiSCo Summary generation.}
We generate DiSCo summaries using the \textit{GPT-5-mini} model\footnote{\url{https://openai.com/index/gpt-5-system-card/}}, guided by a structured prompt designed to capture both expressed and unexpressed aspects of guest experience (see Appendix~\ref{appendix:summary_prompt} for the full system and user prompts). For each accommodation, the LLM is prompted with the seven most frequently mentioned topics, the seven over-represented topics (those with the highest LvS scores), and the seven expected-but-missing topics (those with the lowest LvS scores). To ground the generation process in actual guest feedback, we provide representative review snippets and aggregated sentiment statistics for each topic.
The input data for every topic is organized in a structured format:
\begin{verbatim}
{
"positive": 0,
"negative": 0,
"neutral": 0,
"total": 0,
"mentioned_more_often": 0,
"missing_but_common": 0,
"snippets": []
}
\end{verbatim}
Here, \texttt{positive}, \texttt{negative}, and \texttt{neutral} represent the number of reviews mentioning the topic with each corresponding sentiment, while \texttt{total} denotes the overall number of mentions. The Boolean fields \texttt{mentioned\_more\_often} and \texttt{missing\_but\_common} indicate whether the topic is over-represented or expected-but-missing, respectively. The \texttt{snippets} field contains up to 20 randomly selected review excerpts for the given accommodation, filtered to exclude single-word snippets. This design ensures that the model’s generated summaries are grounded in the original review content, highlighting both frequently expressed aspects and those notably absent, resulting in expectation-aware and contextually balanced representations.

\subsection{Baseline Summaries}
The baseline summaries are designed as a presence-only control that reflects how current LLM-based review summarization systems typically operate. In this condition, the model is prompted exclusively with the most frequently mentioned aspects and representative snippets from the focal accommodation, without any explicit information about expected-but-missing aspects. This design choice allows us to hold the underlying LLM, data source, and prompt structure constant, while isolating the effect of explicitly modeling and surfacing absence through domain-level expectations.

To assess the robustness of the end-to-end generation pipeline in the absence of gold summaries, we conducted a qualitative error analysis of 30 randomly selected summaries across both conditions (DiSCo and Baseline). Two authors manually inspected each summary and verified whether all mentioned aspects and claims could be traced back to the underlying review snippets provided to the model. We define hallucinated content as statements that introduce aspects, attributes, or experiences not supported by the input topic corpus. We did not observe such unsupported content in the inspected summaries. This spot-checking provides additional confidence that the generated summaries remain grounded in the source data, while differing only in whether expectation-based contrasts are explicitly surfaced.


\section{Human Study Methodology}
We conducted a remote study in which 270 participants ranked the summaries based on the criteria we defined. The study involving human participants was reviewed and approved by the [Omitted due to anonimity]. 
Participants were recruited through the Prolific platform, where they reviewed an online consent form and indicated their informed consent prior to participation. All participants received fair monetary compensation and could exit at any time without penalty. No personally identifying information was collected.

\section{Experiment Design}
We hypothesize that summaries differing in their emphasis on distinctive and missing content will lead to measurable differences in how users interpret and evaluate information about accommodations.
In particular, we expect that summaries capturing both salient details and underrepresented aspects will help users form more accurate, detailed, and internally consistent impressions than summaries based solely on dominant, frequently mentioned features.
These differences are expected to reflect not only perceived quality but also how easily readers integrate subtle cues—such as absent amenities or unbalanced opinions—into their overall assessment.
To test this hypothesis, we conducted a comparative user study evaluating whether DiSCo summaries differ significantly from baseline LLM-generated summaries across four quality dimensions: accuracy, detail, consistency, and overall quality, as well as overall user preference.
For each accommodation, two summaries were produced:
(1) a baseline summary, generated directly by an LLM; and
(2) an enhanced summary, generated using DiSCo.
\begin{figure}[h]
\centering
\includegraphics[width=0.9\linewidth]{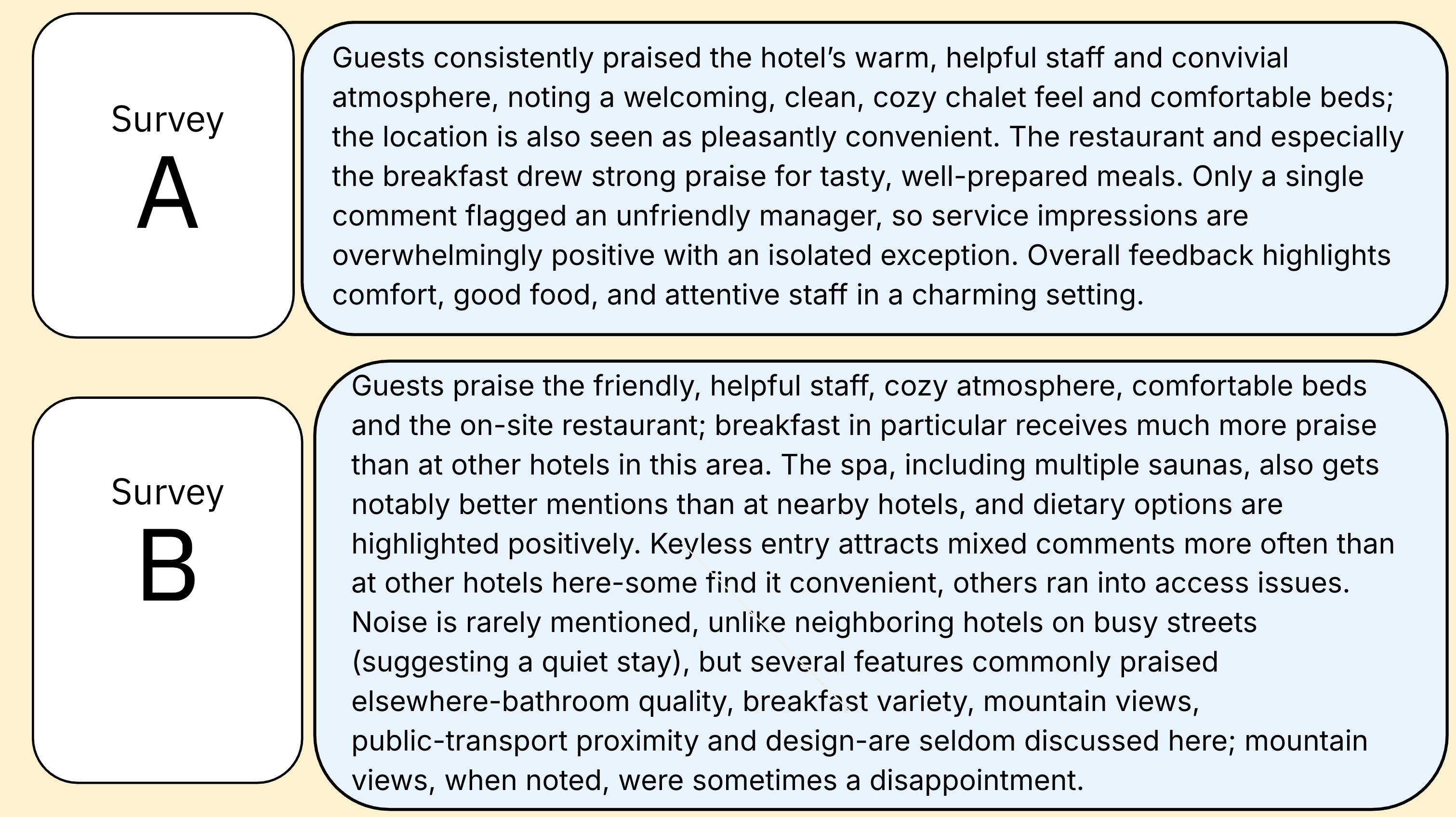}
\caption{\centering Example of a study presenting two versions of a summary for a single accommodation}
\label{fig:sbs}
\end{figure}

\subsection{Participants and Design}
 
Participants were evenly distributed across the three accommodation domains: Ski, Beach, and City-center.   
All participants were fluent in English and reported it as their primary spoken language. 
Demographic characteristics and familiarity levels for each domain are summarized in Table~\ref{tab:demographics_combined}. 
Participants' education level and ethnicity are reported in aggregate form across all conditions.

\begin{table}[h]
\centering
\caption{\centering Participant demographics across accommodation domains.}
\label{tab:demographics_combined}
\begin{tabular}{lcccccc}
\toprule
\textbf{Domain} & \textbf{N} & \textbf{Female} & \textbf{Male} & \textbf{Prefer not to say} & \textbf{Mean age (SD)} & \textbf{Familiarity (\%)} \\
\midrule
Ski          & 90 & 42 & 46 & 2 & 44.6 (12.48) & 25\% \\
Beach        & 90 & 44 & 44 & 2 & 44.9 (12.8)  & 82\% \\
City-center  & 90 & 53 & 37 & 0 & 43.7 (11.4)  & 76\% \\
\midrule
\textbf{Total} & 270 & 139 & 127 & 4 & 44.3 (12.3) & -- \\
\bottomrule
\end{tabular}

\vspace{8pt}
\begin{minipage}[t]{0.49\linewidth}
\centering
\begin{tabular}{lcc}
\toprule
\textbf{Education Level} & \textbf{Count} & \textbf{Percentage} \\
\midrule
High School & 64 & 23.7\% \\
Associate’s Degree & 22 & 8.2\% \\
Bachelor’s Degree & 121 & 44.8\% \\
Master’s Degree & 44 & 16.3\% \\
Doctorate & 6 & 2.2\% \\
Other & 13 & 4.8\% \\
\bottomrule
\end{tabular}
\end{minipage}
\hfill
\begin{minipage}[t]{0.49\linewidth}
\centering
\begin{tabular}{lcc}
\toprule
\textbf{Ethnicity} & \textbf{Count} & \textbf{Percentage} \\
\midrule
White & 215 & 79.6\% \\
Asian & 24 & 8.9\% \\
Black & 15 & 5.6\% \\
Mixed & 15 & 5.6\% \\
Other & 1 & 0.4\% \\
\bottomrule
\end{tabular}
\end{minipage}
\end{table}

Each participant viewed \textbf{18 accommodation summaries} derived from \textbf{9 randomly selected accommodations}, evenly distributed across three categories: \textit{ski resorts}, \textit{beach accommodations}, and \textit{city-center accommodations} (three per category).

For each accommodation, two summaries were generated—one by a baseline LLM and another as a \textit{DiSCo} summary, designed to emphasize information that is missing or notably underrepresented relative to population-level expectations.

Participants were shown a pair of summaries for each accommodation, with the order randomized to avoid sequencing effects. Due to survey platform constraints, all summaries were displayed as high-resolution JPEG images within a \textit{Typeform} survey. The images were generated to preserve full textual fidelity and consistent formatting across devices, and were internally verified to ensure that readability was not affected by image quality or compression.

We compared summaries for three accommodations in each domain—that is, Ski, Beach, and City-center. Each accommodation summary was presented in two versions (baseline and DiSCo) and evaluated by 30 participants, with 15 viewing the baseline first and 15 viewing the DiSCo version first. This design resulted in $3 \times 2 \times 15 = 90$ responses per domain and a total of 270 survey responses overall.

\subsection{Evaluation Process and Metrics}
Participants assessed two alternative summaries describing the same reviewed entity (e.g., a hotel). Each summary differed in the aspects mentioned and the level of detail provided. After reading both versions, participants completed a structured questionnaire composed of five quantitative evaluation items and one overall preference item, followed by two open-ended questions. All responses were recorded in a unified \textit{Typeform} interface using a five-point Likert scale (1 = Not at all, 5 = Very much), with a binary A/B choice for overall preference. All responses were collected anonymously and exported for analysis.

\subsubsection{Process}
Participants first read the two summaries and then rated each along five dimensions---\textbf{Relevance}, \textbf{Detail and Specificity}, \textbf{Decision Support}, \textbf{Persuasive Impact}, and \textbf{Ease of Understanding}.  
These measures together captured complementary facets of perceived summary quality, ranging from informational value to behavioral influence and readability.  
Finally, participants indicated an \textbf{Overall Preference} between the two summaries and explained their reasoning in two open-ended items:  
\textit{Overall, considering all aspects mentioned in each summary, which summary do you prefer?} (binary choice A/B) 
\textit{What are the main reasons for your preference?} (open-ended response)  
This mixed-format design combined quantitative comparability with qualitative insight into participants’ evaluative reasoning.

\subsubsection{Evaluation Metrics}
Participants evaluated each summary along five quantitative dimensions and one overall preference item. Each dimension was defined as follows:

\textbf{Relevance} --- Assesses how meaningful and decision-oriented the mentioned aspects are. A relevant summary emphasizes topics that genuinely matter to travelers (e.g., cleanliness, staff friendliness, Wi-Fi quality) rather than peripheral details such as elevator design.  
\textit{Question: ``How relevant are the aspects mentioned in summary~A for helping you choose your preferred accommodation?''}

\textbf{Detail and Specificity} --- Evaluates the depth and concreteness of the information provided. Detailed summaries include explicit examples or descriptions (e.g., ``the breakfast included fresh fruit and local pastries'') rather than vague generalities (``the breakfast was good'').  
\textit{Question: ``How detailed and specific is the information provided for each aspect in summary~A?''}

\textbf{Decision Support} --- Measures how effectively the presented aspects help users make an informed choice. High-scoring summaries highlight factors that guide real-world booking decisions (e.g., proximity to transport, noise level, service quality).  
\textit{Question: ``To what extent do the aspects presented in summary~A help you make an informed decision about the reviewed entity?''}

\textbf{Persuasive Impact} --- Captures the practical influence of the summary on a participant’s willingness to stay at the accommodation. It reflects the perceived usefulness of the text as a determinant of action.  
\textit{Question: ``How much does summary~A help you decide whether to stay at this accommodation during your trip?''}

\textbf{Ease of Understanding} --- Reflects clarity, coherence, and readability. High-quality summaries are easy to follow, logically structured, and balanced in their presentation of strengths and weaknesses.  
\textit{Question: ``How easy is summary~A to read and understand?''}

\textbf{Overall Preference} --- A binary choice indicating which of the two summaries the participant preferred overall, integrating all preceding dimensions into a holistic comparative judgment.  
\textit{Question: ``Which of the two summaries did you prefer overall?''}

Together, these measures capture both informational quality (Relevance, Detail and Specificity, Decision Support) and experiential quality (Persuasive Impact, Ease of Understanding, Overall Preference). This framework was designed to reveal how \textit{DiSCo} summaries—those highlighting information that is unusually missing or underrepresented relative to comparable accommodations—affect perceived informativeness and decision support compared with standard, presence-focused baselines. In this way, the evaluation aligns directly with the study’s goal of examining how making absences explicit reshapes users’ judgments of summary quality and usefulness.

\section{Results}

\subsection{Evaluation Metrics}
A total of 270 participants rated accommodation summaries under two conditions of \textbf{Baseline} and \textbf{DiSCo}. The five evaluative dimensions were rated using five-point Likert scales (1 = Not at all, 5 = Very much). For each domain, paired-sample \textit{t}-tests compared mean ratings between the two conditions, and effect sizes were calculated using Cohen’s \textit{d} (\textit{.20 = small}, \textit{.50 = moderate}, \textit{.80 = large}). Significance levels are indicated as *\,\textit{p}\,$<$.05, **\,\textit{p}\,$<$.01, ***\,\textit{p}\,$<$.001.

\vspace{0.5em}
\noindent
\textbf{Ski domain.}
As shown in Table~\ref{tab:results_ski}, summaries generated from DiSCo were rated significantly higher than Baseline summaries on all evaluative dimensions except Ease of Understanding, which slightly declined. The largest difference was found for \textit{Detail and Specificity} (\textit{t}(89)=5.62, \textit{p}<.001, \textit{d}=.49), indicating that the inclusion of missing yet expected aspects made the content richer and more informative. Gains in \textit{Helpfulness} and \textit{Decision Support} suggest that participants perceived DiSCo summaries as more actionable.

\begin{table}[h]
\centering
\caption{\centering Ski domain: Mean ratings (Baseline vs. DiSCo), \textit{t}-tests, significance, and effect sizes.}
\label{tab:results_ski}
\begin{tabular}{lcccc}
\toprule
\textbf{Dimension} & \textbf{Baseline (M)} & \textbf{DiSCo (M)} & \textbf{\textit{t}} & \textbf{Cohen’s \textit{d}} \\
\midrule
Relevance & 4.06 & 4.28 & 2.28* & 0.24* \\
Detail and Specificity & 3.71 & 4.10 & 5.62*** & 0.49** \\
Helpfulness & 3.99 & 4.23 & 3.03** & 0.31* \\
Decision Support & 4.10 & 4.29 & 2.47* & 0.26* \\
Ease of Understanding & 4.38 & 4.18 & -2.45* & 0.26* \\
\bottomrule
\end{tabular}
\end{table}

\vspace{0.5em}
\noindent
\textbf{Beach domain.}
For the Beach accommodations (Table~\ref{tab:results_beach}), DiSCo were again evaluated more favorably in most dimensions, with significant increases in \textit{Detail and Specificity} (\textit{t}(89)=6.18, \textit{p}<.001, \textit{d}=.56**) and \textit{Helpfulness} (\textit{t}(89)=3.01, \textit{p}=.006, \textit{d}=.30*). While Relevance showed a small positive trend (\textit{p}=.056), \textit{Decision Support} differences were not significant. Ease of Understanding, however, decreased notably (\textit{t}(89)=-5.30, \textit{p}<.001), indicating that richer content required slightly greater cognitive effort to process.

\begin{table}[h]
\centering
\caption{\centering Beach domain: Mean ratings (Baseline vs. DiSCo), \textit{t}-tests, significance, and effect sizes.}
\label{tab:results_beach}
\begin{tabular}{lcccc}
\toprule
\textbf{Dimension} & \textbf{Baseline (M)} & \textbf{DiSCo (M)} & \textbf{\textit{t}} & \textbf{Cohen’s \textit{d}} \\
\midrule
Relevance & 4.11 & 4.32 & 1.93 & 0.20* \\
Detail and Specificity & 3.54 & 4.17 & 6.18*** & 0.56** \\
Helpfulness & 3.89 & 4.21 & 3.01** & 0.30* \\
Decision Support & 3.98 & 4.11 & 1.17 & 0.12 \\
Ease of Understanding & 4.66 & 4.06 & -5.30*** & 0.55** \\
\bottomrule
\end{tabular}
\end{table}

\vspace{0.5em}
\noindent
\textbf{City-center domain.}
In the City-center results (Table~\ref{tab:results_city}), DiSCo achieved the strongest overall gains, particularly in \textit{Detail and Specificity} (\textit{t}(89)=7.98, \textit{p}<.001, \textit{d}=.84***), representing a large effect size. Improvements were also observed in \textit{Helpfulness} (\textit{t}(89)=2.63, \textit{p}=.010, \textit{d}=.28*) and \textit{Decision Support} (\textit{t}(89)=2.15, \textit{p}=.035, \textit{d}=.23*). Relevance did not differ significantly, and \textit{Ease of Understanding} again decreased, suggesting that participants found Baseline summaries slightly easier to read.

\begin{table}[h]
\centering
\caption{\centering City-center domain: Mean ratings (Baseline vs. DiSCo), \textit{t}-tests, significance, and effect sizes.}
\label{tab:results_city}
\begin{tabular}{lcccc}
\toprule
\textbf{Dimension} & \textbf{Baseline (M)} & \textbf{DiSCo (M)} & \textbf{\textit{t}} & \textbf{Cohen’s \textit{d}} \\
\midrule
Relevance & 4.40 & 4.34 & -0.44 & 0.05 \\
Detail and Specificity & 3.54 & 4.41 & 7.98*** & 0.84*** \\
Helpfulness & 3.96 & 4.30 & 2.63* & 0.28* \\
Decision Support & 4.10 & 4.34 & 2.15* & 0.23* \\
Ease of Understanding & 4.67 & 3.90 & -6.11*** & 0.68** \\
\bottomrule
\end{tabular}
\end{table}

\vspace{0.5em}
\noindent
\textbf{General comparison across domains.}
When aggregating data across all domains (Table~\ref{tab:results_general}), DiSCo were rated significantly higher in \textit{Detail and Specificity} (\textit{t}(269)=8.21, \textit{p}<.001, \textit{d}=.62**), \textit{Helpfulness} (\textit{t}(269)=3.94, \textit{p}<.001, \textit{d}=.29*), and \textit{Decision Support} (\textit{t}(269)=3.22, \textit{p}=.002, \textit{d}=.19). Relevance showed a small but significant improvement (\textit{p}=.05), while Ease of Understanding decreased moderately (\textit{p}<.001, \textit{d}=.51**). The overall pattern thus indicates that DiSCo improved informational and decision-support qualities while modestly increasing reading effort.

\begin{table}[h]
\centering
\caption{\centering Across all domains: Mean ratings, \textit{t}-tests, significance, and effect sizes.}
\label{tab:results_general}
\begin{tabular}{lcccc}
\toprule
\textbf{Dimension} & \textbf{Baseline (M)} & \textbf{DiSCo (M)} & \textbf{\textit{t}} & \textbf{Cohen’s \textit{d}} \\
\midrule
Relevance & 4.19 & 4.31 & 1.99* & 0.12 \\
Detail and Specificity & 3.60 & 4.23 & 8.21*** & 0.62** \\
Helpfulness & 3.94 & 4.25 & 3.94*** & 0.29* \\
Decision Support & 4.06 & 4.25 & 3.22** & 0.19 \\
Ease of Understanding & 4.57 & 4.04 & -6.18*** & 0.51** \\
\bottomrule
\end{tabular}
\end{table}

\vspace{0.5em}
\noindent
\textbf{Interpretation.}
Across all analyses, DiSCo summaries consistently improved evaluative judgments related to informational richness and decision relevance. The strongest effects were found for \textit{Detail and Specificity}, followed by \textit{Helpfulness} and \textit{Decision Support}. These results suggest that incorporating expectation-based absences—i.e., aspects typically present but omitted in the focal entity—enhances perceived informativeness and decision value. The modest decline in \textit{Ease of Understanding} reflects a trade-off between depth and cognitive effort: surprisal-based phrasing invites closer attention and deeper processing without substantially reducing readability.

\subsection{Overall Preference}

Each survey concluded with a forced-choice question asking which of the two versions participants preferred. 
Table~\ref{tab:preference} summarizes participants’ overall preferences between the baseline and DiSCo across the three accommodation domains. 
To test whether the observed proportion of preferences for the DiSCo differed significantly from random (0.5) we used a binomial test within each domain. 
In addition, a chi-square test of independence examined whether the overall preference distribution varied by domain.

\begin{table}[h]
    \centering
    \caption{\centering Summary of preferences by domain. Numbers in parentheses indicate counts out of 90 participants per domain.}
    \begin{tabular}{lccc}
        \toprule
        \textbf{Domain} & \textbf{Baseline (\%)} & \textbf{DiSCo (\%)} & \textbf{Binomial $p$}\\
        \midrule
        Beach        & 50\% (45) & 50\% (45) & 1.00 \\
        City-center  & 47\% (42) & 53\% (48) & 0.56 \\
        Ski          & 30\% (27) & 70\% (63) & $<.001$ \\
        \bottomrule
    \end{tabular}
    \label{tab:preference}
\end{table}

Only the \textit{Ski} domain showed a statistically significant deviation from chance, 
with a clear majority preferring the DiSCo summaries ($70\%$ vs.\ $30\%$, $p<.001$). 
In the \textit{Beach} and \textit{City-center} domains, preferences were approximately balanced ($p=1.00$ and $p=0.56$, respectively). 
A chi-square test of independence comparing the three domains (Beach, City-center, Ski) confirmed that the distribution of preferences was not uniform across conditions, $\chi^2(2)=15.6$, $p<.001$. 

These results suggest that overall preference for DiSCo summaries varies by domain, with the strongest effect observed in the Ski condition. We note that preference is a holistic measure that reflects trade-offs between informativeness and ease of understanding. Although DiSCo summaries consistently improved perceived detail and helpfulness across all domains, their richer content may not translate into a clear preference in settings where brevity or ease of reading is prioritized.

\subsection{Qualitative Insights: Perceived Trade-offs Between Brevity and Completeness}

The qualitative analysis is intended as an exploratory complement to the quantitative results rather than as a comprehensive qualitative study. Our goal is to probe how users cognitively interpret and make use of DiSCo summaries. In particular, we focus on evidence of comparative reasoning, expectation checking, and trade-off assessment, which are central to the proposed role of DiSCo as an expectation scaffolding mechanism.

Each survey concluded with participants stating their overall preference and providing an open-ended response to the question: \textit{``What are the main reasons for your preference?''}.
Here we synthesize the main themes that emerged from these explanations.  

Across domains, participants’ responses revealed a consistent trade-off between brevity and completeness. 
Those who preferred the \textit{baseline summaries} valued their clarity, readability, and focus, describing them as ``shorter and easier to read'', ``clear and coherent'', and ``written in plain English''. 
They felt these versions ``covered the main points well enough'' without digression, and interpreted brevity as a sign of focus and authenticity. 
As one ski participant noted, \textit{``It’s a bit more to the point and easier to read—you get everything you need, some good and also some bad, so you know what to expect if you did stay there''}. 
Similarly, a city-center respondent emphasized that the baseline version ``caught the eye with details necessary for making an informed decision without rambling on about another place''. 
For these participants, a good summary was efficient and human-like, avoiding redundancy while maintaining a natural flow.

In contrast, participants favoring the \textit{DiSCo} emphasized their richness and balance. 
They appreciated the inclusion of both positive and negative aspects, describing them as ``more detailed and honest'', ``a better guide as it mentions negatives as well as positives'', and ``providing a full picture of the accommodation''. 
Detail was often interpreted as a marker of honesty: ``It felt honest and noted things in comparison to what is mentioned elsewhere'', and ``It was explicit in describing the pros and cons without being biased in any way''. 
Across all domains, participants valued that the DiSCo versions surfaced information typically omitted in ordinary reviews—details about accessibility, parking, odors, or noise—that helped them feel better equipped to decide. 
A ski participant remarked, \textit{``It makes it clear that there is no elevator and this indicates, for some, that access may be difficult''}, while a city-center participant appreciated that the review anticipated common urban concerns such as d\'ecor and parking by clarifying that the rooms were freshly renovated and parking was very good. 
On the beach domain, others noted that the richer version ``gave more information and felt more balanced'' or ``mentioned negatives as well as positives'', making it ``a better guide to decide if the accommodation is right for me''. 

Preferences thus reflected participants’ goals and cognitive load: when readability and fluency were prioritized, baseline summaries were favored; when decision confidence and specificity mattered, DiSCo were preferred. 
While this general pattern held across domains, its expression varied. 
Beach participants emphasized completeness and comparative insight; ski participants highlighted honesty and decision relevance, frequently linking detail to action (e.g., ``More information provided with pros and cons so I can make an informed decision''); and city-center participants valued contextual realism and authenticity, noting that the DiSCo felt ``more comparative and less ‘estate agent’ in tone'' and ``like someone actually stayed there, not marketing text''. 

Overall, participants responded positively to the inclusion of explicit absences, suggesting that mentioning what is \textit{not} available enhances perceived informativeness and decision support without diminishing credibility. Though, as one participant observed, not without risks: ``More detailed, including some clues that there are better places elsewhere''.

\section{Discussion}

\subsection{Theoretical Implications}

Across all domains, DiSCo consistently outperformed baseline summaries on dimensions of detail, specificity, and decision support. These results provide empirical support for the role of \textit{expectation-based reasoning} in how people interpret textual summaries. Participants judged summaries as more informative when they reflected not only observed information but also implicit expectations of what \emph{should} appear. This pattern aligns with the well-documented \textit{feature-positive effect}~\cite{newman1980feature,hearst1989backward}, whereby individuals detect present cues more readily than absent ones unless expectations are explicitly invoked. When absence information is surfaced, it appears to activate deeper cognitive processing akin to violation-of-expectation mechanisms~\cite{margoni2024violation}, making invisible gaps perceptible and meaningful. Prior work on the perception of absence shows that missing expected information is rarely detected reliably when left implicit~\cite{shoshan2026makingabsencevisibleroles}. 

At the same time, the observed trade-off between \textit{ease of understanding} and \textit{detail and specificity} may indicate that perceiving absences imposes additional cognitive demands. Detecting what is missing requires not only attention but also comparison with an internal or external reference model. This is a process that may engage analytic, reflective reasoning rather than rapid intuitive judgment. In dual-process terms, absence detection might recruit the slower, deliberative ``System~2'' processes~\cite{kahneman2011thinking}, while presence-based perception relies on fast, automatic ``System~1'' processing. This interpretation suggests an intriguing open question for future research: do individual differences in \textit{need for cognition}~\cite{cacioppo1982need}, the tendency to engage in and enjoy effortful cognitive activity, moderate how users perceive and value absence information? If so, decision-support systems could adjust how much detail or explanation they provide, depending on users’ motivation and willingness to engage in critical thinking.

\subsection{Practical Implications for Intelligent Interfaces and Decision Support}

Beyond summarization, these findings have broader implications for \textit{AI-assisted decision-making systems}, including recommender systems, travel planners, and personalized assistants. Such systems typically optimize for relevance and persuasion, surfacing options that match explicit preferences or frequent patterns in historical data. Yet they rarely reveal what is \emph{not} shown. Items, attributes, or criteria omitted due to bias, sparsity, or model filtering. The present results suggest that making these absences explicit can substantially improve users’ ability to judge completeness, calibrate trust, and make better-informed choices.

In recommender contexts, an \textit{absence-aware} design could highlight what comparable alternatives include but the current recommendation lacks (e.g., ``unlike similar accommodations, this one has no gym'' or ``few users mention accessibility features here''). By externalizing expectations in this way, systems help users reason about trade-offs rather than passively accepting presented options. This aligns with emerging goals in \textit{explainable and value-sensitive AI}, where transparency extends beyond why something was recommended to include why other possibilities were excluded~\cite{friedman2019value,miller2019explanation}.

More generally, absence-aware explanations could augment decision support tools in domains such as healthcare, finance, or education, where missing features (unavailable tests, unreported risks, unaddressed skills) can be as critical as present ones. Presenting absences as structured contrasts rather than negative space supports a shift from \textit{selection} to \textit{reflection}: users can question coverage and infer missing dimensions of evidence. 

At the same time, the observed trade-off between informativeness and ease of understanding highlights the importance of adaptive, interactive presentation. Interactive recommender and decision-support interfaces can employ \textit{progressive disclosure} and \textit{user-contingent interaction}: concise, presence-based summaries can appear by default, while deeper, DiSCo explanations are revealed through user action, such as hovering, clicking, or requesting more detail. Such interactivity enables users to control the cognitive depth of engagement, accessing additional context when motivation or task demands warrant it. In this way, adaptive interaction balances cognitive effort with informational value, ensuring that transparency enhances rather than overloads decision quality.

\subsection{Practical and Societal Implications}
DiSCo summarization has direct implications for transparency, fairness, and inclusion in AI-mediated decision support. By revealing not only what reviews emphasize but also which expected aspects are missing, such systems help users form more complete and reliable judgments. This is particularly valuable for users with accessibility, safety, or ethical concerns who may otherwise be misled by fluent but incomplete AI summaries. Beyond travel recommendations, expectation-based summarization can inform healthcare, e-commerce, and recruitment platforms, where unseen omissions can lead to inequitable or unsafe outcomes. By making informational gaps visible, intelligent interfaces can promote accountability and trust, helping AI systems do good in the world by supporting informed, equitable human decisions.

\subsection{Generalizability Beyond Aspect-Sentiment-Based Review Summarization}
While DiSCo is instantiated in this work using aspect–sentiment analysis of reviews, the underlying principle is not tied to sentiment or opinionated content. At its core, DiSCo operationalizes absence as deviation from an expectation reference: a comparison between an entity’s observed feature distribution and a population-level distribution representing what is typically present in comparable entities. Aspect–sentiment tuples constitute one convenient feature space for review summarization, but the same expectation-based contrast applies to other representations.

For example, in safety analytics, one could summarize accident reports by vehicle type using distributions over accident categories (e.g., collision types, weather conditions, road contexts). A vehicle whose reports lack mentions of specific accident types that are common for comparable vehicles exhibits an informative absence, suggesting relative safety with respect to those risks. No sentiment analysis is required; absences are defined purely over structured event features.

Similarly, in political or policy analysis, speeches can be summarized using word distributions. Comparing a politician’s speeches to those of peers, or to their own past speeches, reveals which expected issues are not addressed. Such absences are often diagnostically meaningful for voters and analysts, even when the language itself is neutral or factual rather than evaluative.

In information-seeking and document search contexts, expectation-based summarization can highlight missing coverage across collections. When summarizing a set of documents returned for a query, the reference distribution may encode typical subtopics or evidentiary dimensions present in the broader corpus. DiSCo summaries can then signal which perspectives, data sources, or criteria are not represented in the retrieved documents, supporting critical evaluation rather than passive consumption.

These examples illustrate that DiSCo generalizes to any domain where (1) users hold structured expectations about what information should be present, and (2) those expectations can be approximated through comparative reference models. Aspect–sentiment analysis is therefore not a limitation of the approach, but a domain-specific instantiation aligned with the review summarization setting studied here.

\subsection{Limitations and Future Work}

This study has several limitations. It assessed subjective perceptions under controlled conditions rather than actual choices, and future work should examine how DiSCo summaries affect real decisions in settings with higher stakes. While the review dataset used in this work is publicly available, LLM-based generation may not be perfectly reproducible despite fixed seeds, which is a general limitation of current generative systems \citep{wang2025assessing}. In addition, aspect–sentiment extraction is used as an intermediate signal rather than a gold-standard annotation, and we rely on lightweight validation rather than exhaustive human labeling. An additional limitation concerns the interpretation of absences themselves. Missing aspects in reviews may reflect reporting norms, cultural practices, review sparsity, or platform-level moderation rather than true deficiencies of an entity. For example, accessibility features may be under-discussed despite being present, or certain negative aspects may be systematically omitted. Accordingly, DiSCo treats absences as expectation-relative signals derived from comparative reference distributions, not as factual claims. Absence-aware summaries should therefore be interpreted as prompts for reflection and comparison rather than definitive assessments. Finally, the hospitality domain offers relatively stable topical patterns, while other domains may require adaptive or alternative reference models.

Future work should explore interactive and multimodal implementations, such as visual or conversational interfaces, that progressively reveal expectation-aware information and allow users to control the level of detail. More generally, these results point toward a broader paradigm of \textit{expectation-aware AI}~\cite{mumford2021absence,kahneman2011thinking}, in which systems represent both what is present and what is expected yet absent, aligning computational surprisal with human interpretive cognition.

\section{Conclusion}
This work demonstrates that making absences explicit in review summarization improves perceived informativeness and decision support. By comparing instance content with domain-level expectations, DiSCo summaries help users recognize missing but relevant information that standard presence-based models overlook. The results highlight a cognitive trade-off: greater insight at the cost of slightly higher effort. Designing adaptive, interactive interfaces that surface absences selectively can balance this trade-off, enabling users to choose when to engage in deeper reasoning. More broadly, these findings point toward expectation-aware AI systems that support decisions not only by presenting what is visible, but also by revealing what is notably absent.

\section*{Generative AI Usage Disclosure}
Generative AI tools -- ChatGPT (GPT-5) utilized exclusively for language refinement of this manuscript. These tools assisted in enhancing grammatical accuracy, textual clarity, and overall readability. All content creation, data analysis, programming, and methodological decisions were conducted without AI assistance. In addition, this study involved the creation and controlled use of custom LLM-based tools as part of the experimental procedure (e.g., for creating summaries). These tools were integral to the research design and are fully described in the Methods section. The authors retain complete responsibility for the scholarly content and intellectual contributions presented in this work.

\bibliographystyle{ACM-Reference-Format}
\bibliography{cites}

\appendix
\section{List of Topics}
\label{appendix:topics}

This appendix presents the full taxonomy of accommodation review topics used in our analysis.

\begin{description}
  \item[\textbf{Room}] cleanliness, size, layout, bed\_comfort, mattress\_quality, pillow\_quality, bathroom\_quality, shower\_water\_pressure, toiletries, noise\_levels, soundproofing, temperature\_control, air\_conditioning, heating, views, in\_room\_facilities, lighting, furniture\_condition, balcony\_or\_terrace.

  \item[\textbf{Cleanliness}] overall\_hotel\_cleanliness, smells, linens\_freshness, towels\_freshness, maintenance, pest\_control.

  \item[\textbf{Service}] staff\_friendliness, staff\_helpfulness, professionalism, multilingual\_staff, concierge\_service, housekeeping\_service, responsiveness\_to\_requests.

  \item[\textbf{Food \& Beverage}] breakfast\_quality, breakfast\_variety, restaurant\_quality, restaurant\_price, room\_service, bar\_quality, dietary\_options, coffee\_quality, local\_cuisine\_availability.

  \item[\textbf{Location}] proximity\_city\_center, proximity\_attractions, proximity\_public\_transport, proximity\_airport, safety\_area, neighborhood\_vibrancy, proximity\_restaurants, proximity\_shops, proximity\_nightlife, proximity\_sea, proximity\_beach, proximity\_ski\_resort, proximity\_hiking\_trails, scenic\_surroundings, mountain\_views, lake\_views.

  \item[\textbf{Value}] value\_for\_money, unexpected\_fees, discounts\_offers, resort\_fees, parking\_fees.

  \item[\textbf{Pool}] pool\_cleanliness, pool\_size, pool\_temperature, pool\_opening\_hours, sunbeds\_availability, pool\_bar.

  \item[\textbf{Gym}] equipment\_quality, equipment\_variety, gym\_cleanliness, gym\_size, opening\_hours.

  \item[\textbf{Spa}] spa\_quality, spa\_variety, massage\_services, wellness\_treatments, sauna, steam\_room.

  \item[\textbf{Business}] business\_center\_quality, meeting\_room\_facilities, conference\_services, printing\_services, wifi\_reliability.

  \item[\textbf{WiFi}] speed, reliability, coverage, ease\_of\_access.

  \item[\textbf{Parking}] availability, price, security, proximity\_to\_hotel.

  \item[\textbf{Accessibility}] wheelchair\_accessibility, elevator\_availability, accessible\_rooms, staff\_assistance.

  \item[\textbf{Family}] family\_room\_availability, crib\_availability, play\_area, babysitting\_service, kids\_club, activities\_for\_families.

  \item[\textbf{Pets}] pet\_policy, pet\_fees, pet\_amenities, dog\_friendly\_areas.

  \item[\textbf{Check-in/Check-out}] speed, flexibility, early\_checkin, late\_checkout, ease\_booking\_payment, keyless\_entry.

  \item[\textbf{Transport}] airport\_transfer, shuttle\_service, taxi\_availability, ride\_share\_access, bike\_rental, ski\_shuttle.

  \item[\textbf{Ski}] proximity\_ski\_lift, ski\_storage, ski\_rental, ski\_in\_ski\_out, drying\_room, ski\_pass\_services.

  \item[\textbf{Entertainment}] live\_music, evening\_shows, casino, karaoke, themed\_events.

  \item[\textbf{Sustainability}] eco\_friendly\_practices, recycling\_program, energy\_efficiency, water\_saving\_measures.

  \item[\textbf{Other}] overall\_atmosphere, design\_style, renovation\_status, special\_experiences, holiday\_decorations.
\end{description}

\section{Aspect Sentiment Analysis Prompts}
\label{appendix:absa_prompts}

\subsection{System Prompt}
\begin{verbatim}
You are an aspect-based sentiment analyzer tasked with extracting sentiment 
(positive, negative, or neutral) for specific aspects from user reviews.

Begin with a concise checklist (3--7 bullets) of what you will do; 
keep items conceptual, not implementation-level.

Each input review is a JSON object containing:
- 'title': the review title.
- 'liked': what the guest liked about the accommodation.
- 'disliked': what the guest disliked about the accommodation.

Note:
- Positive points may occasionally appear in the 'disliked' section, 
  and negative points in the 'liked' section.
- Map all aspects to the closest match from the provided aspect leaf list 
  (aspect leaves are grouped under parent categories) if either an exact match 
  or a close paraphrase/typo is found; otherwise, ignore the mention.

For each review, output a JSON array. Each array element should be an object 
with the following keys:
- aspect_parent: the parent category (e.g., "CheckIn_CheckOut").
- aspect_leaf: the specific aspect leaf, as defined in the provided list 
  (e.g., "ease_booking_payment").
- sentiment: one of "positive", "negative", or "neutral".
- snippet: the snippet from the review that includes the aspect and sentiment 
  (e.g., "breakfast was great").

Guidelines:
- If a review contains multiple or conflicting sentiments for the same aspect, 
  list each sentiment separately along with its corresponding aspect.
- Maintain the order of aspect mentions as they appear in the review.
- Only include aspects from the provided list; discard others.
- If no valid aspects are found, output an empty array.

After forming the output, briefly verify that all included aspects are from the 
provided list and that sentiment labels are consistent with the extracted 
statements. If any inconsistency is found, correct the output.

Output Format:
[
  {
    "aspect_parent": "Room",
    "aspect_leaf": "cleanliness",
    "sentiment": "positive",
    "snippet": "room was very clean"
  },
  {
    "aspect_parent": "Service",
    "aspect_leaf": "staff_friendliness",
    "sentiment": "negative",
    "snippet": "receptionist was rude"
  },
  {
    "aspect_parent": "Room",
    "aspect_leaf": "cleanliness",
    "sentiment": "negative",
    "snippet": "There was a used towel in bathroom"
  }
]

Example output with no aspects:
[]
\end{verbatim}

\subsection{User Prompt}
\begin{verbatim}
Aspects: {aspects}
Review: {review}
\end{verbatim}

\section{Summary Generation Prompts}
\label{appendix:summary_prompt}

\subsection{System Prompt}
\begin{verbatim}
You are an expert at summarizing accommodation guest feedback.
You are given topic-level aggregated data for the accommodation.
Each topic entry includes:
- topic: the theme of guest feedback
- sentiment_counts: {positive, negative, neutral}
- mentioned_more_often: true/false  (significantly more mentions than the reference group)
- missing_but_common: true/false    (common in reference group but rarely mentioned here)
- sample_snippets: optional example phrases (for context only)

*Reference group for comparison*: similar accommodations in the area

Your task:
Write a *fluent, human-readable summary* for potential guests.

*Guidelines*
1. *Summarize mentioned topics* with their sentiment (positive/negative/neutral)

2. *Highlight unusual emphasis*: When "mentioned_more_often" is true, explicitly compare to the reference group:
   - Good: "The rooftop pool gets much more praise than at other hotels in this district"
   - Avoid: "The pool is mentioned more often" (too vague)

3. *Interpret absences with context* when "missing_but_common" is true:

   *Positive absence*:
   - "Noise is rarely mentioned, unlike neighboring hotels on this busy street—suggesting better soundproofing"

   *Neutral absence*:
   - "Breakfast isn't discussed, unlike most hotels in this area—these are self-catering apartments"

   *Concerning absence*:
   - "Staff friendliness isn't mentioned, though it's commonly praised at similar hotels nearby"

4. *Always specify the comparison* when noting differences:
   - Use: "other hotels in [area]", "similar [type] hotels", "nearby hotels"
   - Don't say: "similar hotels", "comparable properties" (too vague)

5. *Do NOT*:
   - List raw numbers or percentages
   - Quote snippets verbatim
   - Make comparisons without specifying to what

6. *Format*: Single cohesive paragraph (max 120 words)
\end{verbatim}

\subsection{User Prompt}
\begin{verbatim}
### Most mentioned topics data:
{most_mentioned_topics}

### Topics that are mentioned more often compared to similar accommodations:
{mentioned_more_often}

### Topics that usually mentioned with a positive sentiment for other accommodations but are under-represented for 
this accommodation: {missing_but_common_positives}

### Topics that usually mentioned with a negative sentiment for other accommodations but are under-represented for
this accommodation:
{missing_but_common_negatives}
\end{verbatim}

\end{document}